\documentclass{PoS}

\usepackage{slashed}
\usepackage[T1]{fontenc}
\usepackage[utf8]{inputenc}
\usepackage{amsmath} 
\usepackage{latexsym}
\usepackage{color}
\usepackage{bm}
\usepackage{amsthm}
\usepackage{amsmath}
\usepackage{amssymb}
\usepackage{graphicx}
\usepackage{mathtools}
\usepackage{prettyref}
\usepackage{comment}

\usepackage[hyphens]{url}

\newtheorem{defi}{Definition}

\title{Matrix norms and search for sterile neutrinos}

\ShortTitle{Matrix norms, neutrino mixings}

\author{\speaker{Wojciech Flieger}, Franciszek Pindel, Kamil Porwit\\ \\
Institute of Physics, University of Silesia, Katowice, Poland 

\\ \\
E-mails: \email{woj.flieger@gmail.com}, \email{franciszek.pin@gmail.com},  \email{kamil.porwit@smcebi.edu.pl}  
}

\abstract{Matrix norms can be used to measure the "distance" between two matrices which translates naturally to the problem of calculating the unitary deviation of the neutrino mixing matrices. Variety of matrix norms opens a possibility to measure such deviations on different structural levels of the mixing matrix. 
}

\FullConference{Corfu Summer Institute 2018 "School and Workshops on Elementary Particle Physics and Gravity"\\
		(CORFU2018)\\
		31 August - 28 September, 2018\\
		Corfu, Greece}

\begin{document}

\section{Introduction}

Neutrino experiments do not exclude that additional right-handed sterile neutrinos exist. 
There are many ways, inspired by theory, in which they can be discovered both in neutrino oscillations \cite{PhysRevD.70.073004, PhysRevD.80.073001, PhysRevLett.107.091801, Abazajian:2012ys,Dib:2019ztn, Gariazzo2017, Gandhi2015} and collider physics \cite{Gluza:2015goa,Gluza:2016qqv,Golling:2016gvc,Dube:2017jgo,Mangano:2018mur,Antusch:2018bgr}. Their masses are practically not limited, ranging from eV to TeV, or the Planck scale. Experiments and theoretical studies are devoted to search for such signals.
Interestingly, some present experiments and the experimental signals suggest that the fourth type of neutrino may exist.

Namely, measurements of the $\bar{\nu}_e$ flux at small distances from nuclear reactors gives 6\% less events than expected \cite{Mueller:2011nm}. Such a deficit referenced as a "Reactor Antineutrino Anomaly" can be explained as active-sterile antineutrino oscillations at very short baselines \cite{PhysRevD.83.073006}. Moreover, the NEUTRINO-4 reactor experiment claims to detect an electron antineutrino to sterile neutrino oscillation at the $3\sigma$ significance level \cite{Serebrov:2018vdw}. Most probable values for oscillation parameters are estimated as sin$^2\theta_{14} = 0.39$ and $\Delta m^2_{14} = 7.34$ eV$^2$.\\ Gallium solar neutrinos experiments also observe less events than predicted. Deficit in data is reported by GALEX and SAGE collaborations \cite{Abdurashitov:2005tb}. Statistical significance for such "Gallium Anomalies" in terms of neutrino oscillations was recently estimated at the level of 3$\sigma$ \cite{Giunti:2010zu}.\\
Clues for sterile neutrino oscillations exist also in the long baseline experiments. First reported abnormalities were reported by the LSND collaboration in 1995. More efficient analysis from 1996 showed an excess of electron antineutrinos events from the muon antineutrino beam \cite{Athanassopoulos:1996jb}. However, some publications cast in doubt analysis done within the LSND and its interpretation in favour of sterile neutrinos. Reduction of the significance level below 3$\sigma$ is claimed \cite{Bolshakova:2011ib, Maltoni:2002xd}. 
Recent results from the MiniBooNE experiment show a massive excess of electron (anti)neutrino events from the muon (anti)neutrino beam. Surplus of (79)281 events in correspondence to predicted number of events was reported \cite{Aguilar-Arevalo:2018gpe}.

However, discrepancies between predicted and observed data events can be explained not only by adding new neutrinos to the existing theory. There are still many exciting possibilities such as: resonant neutrino oscillations \cite{Karagiorgi:2012kw, Doring:2018cob, PhysRevD.97.075021}, Lorentz violation \cite{PhysRevD.69.016005, PhysRevD.85.016013}, sterile neutrino decay \cite{Bertuzzo:2018itn, Ballett:2018ynz} and many more.

Here we will focus on the theoretical analysis connected with deviations from unitarity of the standard mixing matrix $U_{PMNS}$. Such deviations would be a clear signal of known neutrinos mixing with additional right-handed sterile states. We extend this kind of analysis and study how deviations from unitarity can be defined in the framework of advanced matrix algebra using matrix norms. 
The advantage of using norms instead of the maximal absolute value of the matrix is of mathematical nature - matrix norms in a natural way reflects properties of matrices. Moreover, they posses important properties which help with theoretical calculations and manipulation of matrices. 
Most of the background for such analysis was established in \cite{Bielas:2017lok} where a notion of singular values, matrix norms and contraction was introduced. This machinery allows us to determine when a non-unitary mixing matrix can describe physical phenomenon. Moreover, it is possible to define a set of all physically admissible mixing matrices. On top of that the method of so-called unitary dilation gives us the possibility to study in a systematic way scenarios with additional neutrinos \cite{Bielas:2017exa}.

\section{Parametrizations of the neutrino mixing matrix}
 
To describe neutrino non-standard mixings usually two parameterizations are used, commonly denoted by $\alpha$ and $\eta$ \cite{FernandezMartinez:2007ms, Antusch:2006vwa, Xing:2008, Xing:2011ur}, defined in the following way
\begin{eqnarray}
U_{PMNS}&=&(I-\eta)V, \nonumber \\
U_{PMNS}&=&(I-\alpha)W.
\end{eqnarray}
Here $\eta$ is a Hermitian matrix, $\alpha$ is a lower triangular matrix and $U,V$ and $W$ are unitary matrices. The size of the deviation is described by matrices $\alpha$ and $\eta$. In the limit $\eta,\alpha \to 0$, the standard $U_{PMNS}$ unitary matrix is restored \cite{Maki:1962mu, Kobayashi:1973fv, Bilenky:1987ty}. 
In contrast to these approaches we want to find an optimal unitarity deviation one-parameter indicator. This value will be estimated using matrix norms which have the property of functions to measure the "distance" between two matrices.
We will show an example of differences between norms when applied to the neutrino data.
 
\section{Matrix norms and deviations from unitarity}

Let us recall the general definition of the matrix norm. We will use it to study a deviation from unitarity of the mixing matrix. 
\begin{defi}
A matrix norm is a function $\Vert \cdot \Vert$ from the set of all complex matrices into $\mathbb{R}$ which for any $A,B \in \mathbb{C}^{n \times n}$ and any $\alpha \in \mathbb{C}$ satisfies the following conditions:
\begin{enumerate}
\item $\Vert A \Vert \geq 0 \quad and \quad \Vert A \Vert =0 \Leftrightarrow A=0 $
\item  $\Vert \alpha A \Vert = \vert \alpha \vert \Vert A \Vert$
\item  $\Vert A + B \Vert \leq  \Vert A \Vert +  \Vert B \Vert$ 
\item  $\Vert AB \Vert \leq \Vert A \Vert \Vert B \Vert $
\end{enumerate} 
\end{defi}
As in the case of the "normal" vector norms there are plenty of matrix norms, each useful for a different application. We will not discuss here all of them but it is important to mention that all of them are equivalent which means that for a given matrix $A \in \mathbb{C}^{n \times m}$ any two matrix norms $\Vert \cdot \Vert_{a}$ and $\Vert \cdot \Vert_{b}$ satisfy
\begin{equation}
k \Vert A \Vert_{a} \leq \Vert A \Vert_{b} \leq s \Vert A \Vert_{a}
\end{equation}
where $k,l$ are positive numbers.

Specific matrix norms are used in many different applications, especially in data analysis, economics, optimization and image processing \cite{Steinberg05computationof, NIPS2010_3988, le:hal-00975276, AGUIAR2017163}.

Some representative examples of the matrix norms are the following
\begin{align}
&\textrm{1. Operator norm (ON): } \Vert A \Vert_{2}=max_{\Vert x \Vert_{2}=1} \Vert A x \Vert _{2} = \sigma_{max}(A) . \label{On} \\
&\textrm{2. Frobenius norm (FN): } \Vert A \Vert_{F}=\sqrt{Trace(A^{\dag}A)}=\sqrt{\sum_{i=1}^{n}\sum_{j=1}^{m}\vert a_{ij} \vert^{2}}=\sqrt{\sum_{i=1}^{min \lbrace n,m \rbrace } \sigma_{i}(A)^{2}}. \label{Fn} \\
&\textrm{3. Maximum absolute column sum norm (MACN): } \Vert A \Vert_{1}= \max_{j} \sum_{i=1}^{m} \vert a_{ij} \vert . \label{Cn} \\
&\textrm{4. Maximum absolute row sum norm (MARN): } \Vert A \Vert_{\infty}= \max_{i} \sum_{j=1}^{m} \vert a_{ij} \vert . \label{Rn}
\end{align}

\noindent Now we are ready to define the way to measure the deviation from unitarity of the mixing matrix.
\begin{defi}
Let $A \in \mathbb{C}^{n \times n}$ be a given matrix and let $\Vert \cdot \Vert$ be any matrix norm then the function
\begin{equation}
\Vert I-AA^{\dag} \Vert \label{def2}
\end{equation}
measures how far $A$ is from the unitary matrix.
\end{defi}

\section{Structure of the physical region for neutrino mixing matrices}

Not all matrices that deviate from unitarity can describe physical mixing. Whether a matrix can be treated as physical is determined by the probability of transition between flavour states. In oscillations we cannot lose neutrinos which means that the probability of transition from a given state to any other must be one. This is ensured by the unitarity of the mixing matrix. Thus physically admissible mixing matrices must be either unitary or can be extended to the unitary matrix. Such matrices are known as contractions, i.e., the matrices with the largest singular value less or equal to one, symbolically $\sigma_{max}(A) \leq 1$. On the other hand we have bounds for mixing parameters obtained from experiments. These two requirements together give us the definition of the physical region. 
A region of physically admissible mixing matrices is defined as a convex hull spanned on all 3-dimensional unitary matrices with parameters bounded by experiments or equivalently as a set of all convex combinations of these unitary matrices. Thus the formal definition looks in the following way
\begin{equation}
\begin{split}
\Omega= \lbrace \sum_{i=1}^{M} \alpha_{i} U_{i} \vert U_{i} \in U_{3 \times 3}, \alpha_{1},...,\alpha_{M} \geq 0, \sum_{i=1}^{M} \alpha_{i}=1, \\
\theta_{12}, \theta_{13}, \theta_{23} \ and \ \delta \ restricted \ by \ experiments \rbrace .
\end{split}
\label{omega}
\end{equation}
It can also be seen as an intersection of the unit ball with respect to the Operator norm with the hyperrectangle determined by experimental ranges. The formal definition of the physical region $\Omega$ opens a possibility for further interesting studies and better understanding of the neutrino mixing.  The first issue is the shape of the physical region. Since the mixing matrices are 9-dimensional objects (real or complex) the exact shape of $\Omega$ is not possible to visualize. However we could gain a great chunk of information by projecting this region onto two or three dimensional subspaces. This problem can be solved by incorporating methods of Semidefinite Programming (SDP) \cite{2009Sanyal, 2014Saunderson}. 
The next issue concerns possible extensions of the Standard Model by introducing sterile neutrinos. It is known that the number of singular values strictly less than one controls the smallest possible unitary extension of the 3-dimensional mixing matrix, i.e. the number of additional neutrinos \cite{Bielas:2017lok}. Thus the collection of matrices that allow the smallest unitary dilation with one, two or three sterile neutrinos divides the region $\Omega$ into three disjoint subsets. It is interesting how these subsets are distributed within $\Omega$. Are they distributed randomly or they clustered? Visualization of the physical region mentioned before can also be helpful. The next important issue is the entrywise characterization of contractions. This will allow to construct matrices from $\Omega$ in a simple way and give entrywise description of this region. There are  more open issues about the region of physically mixing matrices like: what is the minimal number of unitary matrices necessary to cover the whole region.

\section{A numerical example}
In what follows, let us construct a matrix from the region $\Omega$ as a convex combination of unitary $U_{PMNS}$ matrices and determine its deviation from unitarity using different norms introduced before.
All calculations have been done with 16 digits precision. However, we will keep 4 significant digits for all numbers given, as an accuracy which can be achieved for error estimation (0.003) in such analysis is at the same level \cite{Bielas:2017lok}.
The matrix $V$ from the physical region $\Omega$ is constructed as the convex combination of 3 unitary $U_{PMNS}$ matrices characterized by the following sets of mixing parameters: 
\begin{equation}
\begin{split}
&U_{1}:
\theta_{12}=0.5681, \ \theta_{13}=0.1425, \ \ \theta_{23}= 0.7277, \\
&U_{2}:
\theta_{12}=0.5973, \ \ \theta_{13}=0.1539, \ \ \theta_{23}= 0.6998,  \\
&U_{3}:
\theta_{12}=0.6279, \ \ \theta_{13}=0.1429,  \ \ \theta_{23}= 0.9219.
\end{split}
\end{equation}
Thus the matrix $V$ is the following sum
\begin{equation}
V=\alpha_{1} U_{1} +\alpha_{2} U_{2} +\alpha_{2} U_{2}, 
\label{constr1}
\end{equation}
where the coefficients are  
\begin{equation}
\alpha_{1}=0.315, \quad
\alpha_{2}=0.313, \quad
\alpha_{3}=0.372. 
\end{equation}
Such a convex combination gives the following mixing matrix
\begin{equation}
V=
\begin{pmatrix}
0.8166 &  0.5580 &  0.1457 \\
-0.4781 &  0.5200  &  0.7000 \\
0.3158 & -0.6406 &  0.6919
\end{pmatrix}.
\end{equation}
Let us first check if this matrix is a contraction. To do this we calculate a set of its singular values
\begin{equation}
\sigma(V)= \lbrace 0.999, 0.994, 0.994 \rbrace .   
\end{equation}
The largest singular value is less than one, therefore the matrix $V$ is a contraction. Although the result can suggest that taking an error into account the $\sigma_{1}(V)$ can possible take the value above one, this is not the case since the construction (\ref{omega},\ref{constr1}) impose the contraction property.
Now let us study deviation from unitarity for this matrix using functions introduced in the Definition 2.
\begin{eqnarray}
\Vert I-VV^{\dag} \Vert_{2}&=& 0.0114, \nonumber \\
\Vert I-VV^{\dag} \Vert_{F}&=& 0.0160, \nonumber \\
\Vert I-VV^{\dag} \Vert_{1}&=& 0.0131, \nonumber \\
\Vert I-VV^{\dag} \Vert_{\infty}&=& 0.0131, \nonumber \\
max \vert I-VV^{\dag} \vert&=& 0.0111.
\end{eqnarray}

We can see that  for the considered example, besides MACN: $\Vert \cdot \Vert_{1}$ and MARN: $\Vert \cdot \Vert_{\infty}$ defined in (\ref{Cn}, \ref{Rn}), other norms give different results. MACN and MARN norms give the same result because we calculate unitarity deviations for the symmetric matrix $VV^{\dag}$. 
The definitions (\ref{Cn}, \ref{Rn}) suggest that these two specific norms can be used to measure in a simple way the largest unitary deviation among mixings with particular flavor or massive states. Unfortunately, this is not the case because in the product $VV^{\dag}$ these states are mixed. However, information about unitarity deviations on the level of flavors states is gathered by the diagonal elements of $VV^{\dag}$. Similarly, information about unitarity deviations on the level of massive states is given by the diagonal elements of $V^{\dag}V$. It is necessary to check whether these elements satisfy requirements of the matrix norm. If not, they are still attractive functions to use in the case of neutrinos as the measure of the unitarity deviation on the levels of flavour and massive states but additional numerical and mathematical studies are necessary.  
If we are interested in the absolute deviation from unitarity the Frobenius norm should be the proper choice as it takes the whole structure of the matrix into account.

\section{Summary}
Matrix theory and matrix analysis are fruitful areas of mathematics with a broad range of applications. They can also be used to give new insights into neutrino mixings analysis \cite{Bielas:2017lok}. 
Here we have discussed a notion of matrix norms and their application in the calculation of deviation from unitarity of the neutrino mixing matrix.
In general, different norms give different numerical estimates. We have made a rough estimate which one would be optimal for the unitary deviation of the neutrino mixing matrix. 
The example shows that the Frobenius norm gives the largest numerical value of the deviation. This agrees with the definition \eqref{Fn} because it takes into account all entries of the matrix. That is why the Frobenius norm seems to be a right choice to be taken if the overall amount of the unitarity deviation of the mixing matrix is discussed. 
The usual way of calculating the unitary deviation using the maximum of absolute value is very close to the result given by the Operator norm \eqref{On}.
 In the current form of the Definition 2 \eqref{def2}  the MACN and MARN norms (\ref{Cn}, \ref{Rn}) do not give expected information.

The ultimate goal in the issue of determining the deviation from unitarity   would be a method to compute it just from the initial matrix without necessity of its Hermitian product $VV^{\dag}$, which is closely related to the problem of the entrywise characterization of contractions. This together with other issues brought up by us should result in a deeper understanding of the structure of the physical region $\Omega$. 
\vspace{0.5cm}

\section*{Acknowledgements}
We would like to thank Janusz Gluza and Marek Gluza for useful remarks and careful reading of the manuscript. 
Work is supported in part by the COST Action CA16201 and the Polish National Science Centre (NCN) under the Grant Agreement 2017/25/B/ST2/01987. {\it F.P.} is supported by the University of Silesia program "Uniwersytet dla Najlepszych" 2018/2019.

\providecommand{\href}[2]{#2}
\addcontentsline{toc}{section}{References}

\bibliographystyle{elsarticle-num}
\bibliography{ref}


\end{document}